\documentclass[ twocolumn,superscriptaddress,tightenlines]{revtex4-1}
\usepackage{amsthm}
\usepackage{amsmath}
\usepackage{graphicx}
\usepackage[usenames,dvipsnames]{color}
\usepackage[colorlinks=true,citecolor=magenta,urlcolor=blue]{hyperref}
\usepackage{colortbl}
\usepackage{color}
\usepackage{multirow}
\usepackage{hhline}
\usepackage{epstopdf}

\newcolumntype{C}[1]{ >{ \centering\arraybackslash}p{#1}}

\newcommand{\X}{$\mathsf{X}$}
\newcommand{\Z}{$\mathsf{Z}$}

\newcommand{\pax}{p_\mathsf{X}^\text{A}}
\newcommand{\paz}{p_\mathsf{Z}^\text{A}}
\newcommand{\pbx}{p_\mathsf{X}^\text{B}}
\newcommand{\pbz}{p_\mathsf{Z}^\text{B}}

\newcommand{\QZ}{Q_\mathsf{Z}}

\newcommand{\sz}[1]{s_{\mathsf{Z}, #1}}

\newcommand{\phiz}{\phi_{\mathsf{Z}}}

\newcommand{\be}{\begin{equation}}
\newcommand{\ee}{\end{equation}}

\newcommand{\del}[1]{}

\newcommand{\sket}[1]{{\ensuremath{\lvert#1\rangle}}}
\newcommand{\lket}[1]{{\ensuremath{\left\lvert#1\right\rangle}}}
\newcommand{\ket}[1]{\if@display\lket{#1}\else\sket{#1}\fi}

\newcommand{\sbra}[1]{{\ensuremath{\langle#1\rvert}}}
\newcommand{\lbra}[1]{{\ensuremath{\left\langle#1\right\rvert}}}
\newcommand{\bra}[1]{\if@display\lbra{#1}\else\sbra{#1}\fi}

\newcommand{\sbraket}[2]{{\ensuremath{\langle#1\rvert#2\rangle}}}
\newcommand{\lbraket}[2]{{\ensuremath{\left\langle#1\!\left\rvert\vphantom{#1}#2\right.\!\right\rangle}}}
\newcommand{\braket}[2]{\if@display\lbraket{#1}{#2}\else\sbraket{#1}{#2}\fi}

\newcommand{\sketbra}[2]{{\ensuremath{\lvert #1\rangle\!\langle #2\rvert}}}
\newcommand{\lketbra}[2]{{\ensuremath{\left\lvert #1\right\rangle\!\!\left\langle #2\right\rvert}}}
\newcommand{\ketbra}[2]{\if@display\lketbra{#1}{#2}\else\sketbra{#1}{#2}\fi}

\begin{document}
\title{Simple 2.5\,GHz time-bin quantum key distribution}

\author{Alberto Boaron}\email{alberto.boaron@unige.ch}
\affiliation{Group of Applied Physics, University of Geneva, Chemin de Pinchat 22, CH-1211 Geneva 4, Switzerland}
\author{Boris Korzh}
\affiliation{Group of Applied Physics, University of Geneva, Chemin de Pinchat 22, CH-1211 Geneva 4, Switzerland}
\author{Raphael Houlmann}
\affiliation{Group of Applied Physics, University of Geneva, Chemin de Pinchat 22, CH-1211 Geneva 4, Switzerland}
\author{Gianluca Boso}
\affiliation{Group of Applied Physics, University of Geneva, Chemin de Pinchat 22, CH-1211 Geneva 4, Switzerland}
\author{Davide Rusca}
\affiliation{Group of Applied Physics, University of Geneva, Chemin de Pinchat 22, CH-1211 Geneva 4, Switzerland}
\author{Stuart Gray}
\affiliation{Corning Incorporated, Corning, NY 14831, United States}
\author{Ming-Jun Li}
\affiliation{Corning Incorporated, Corning, NY 14831, United States}
\author{Daniel Nolan}
\affiliation{Corning Incorporated, Corning, NY 14831, United States}
\author{Anthony Martin}
\affiliation{Group of Applied Physics, University of Geneva, Chemin de Pinchat 22, CH-1211 Geneva 4, Switzerland}
\author{Hugo Zbinden}
\affiliation{Group of Applied Physics, University of Geneva, Chemin de Pinchat 22, CH-1211 Geneva 4, Switzerland}

\begin{abstract}
We present a 2.5\,GHz quantum key distribution setup with the emphasis on a simple experimental realization. It features a three-state time-bin protocol based on a pulsed diode laser and a single intensity modulator. Implementing an efficient one-decoy scheme and finite-key analysis, we achieve record breaking secret key rates of 1.5\,kbps over 200\,km of standard optical fiber.  
\end{abstract}

\maketitle

Since the invention of quantum key distribution (QKD) in 1984~\cite{Bennett1984} and the first experimental realizations in the 1990’s \cite{Gisin2002}, the progress has been constant, and first commercial devices are available and used for secure communication.
State-of-the-art academic experiments feature GHz pulse rates~\cite{Takesue2007, Wang2012, Shibata2014, Dynes2016, Grunenfelder2018}, MHz secret bit rates over short distances~\cite{Dynes2016} and a reach of more than 300\,km of low loss fibers \cite{Korzh2015, Yin2016, Shibata2014}.
Moreover, recently QKD with satellites has been demonstrated \cite{Liao2017}.
In this paper, we are pushing QKD further to its limits, presenting a system with 2.5\,GHz pulse rate which is at the same time simple and efficient, improving the secret key rate (SKR) by one order of magnitude at 200\,km.

\begin{figure}[b]
\begin{tabular}{>{\centering\arraybackslash}m{2cm}|>{\centering\arraybackslash}m{2cm}|>{\centering\arraybackslash}m{2cm}|>{\centering\arraybackslash}m{2cm}}
basis, bit&state&$\mu_1$&$\mu_2$\\
\hline
\Z{}, 0&$\ket{\psi_0}$&\includegraphics[width=1cm]{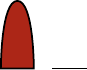}&\includegraphics[width=1cm]{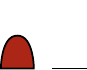}\\[\smallskipamount]
\Z{}, 1&$\ket{\psi_1}$&\includegraphics[width=1cm]{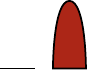}&\includegraphics[width=1cm]{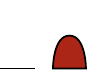}\\[\smallskipamount]
\hline
\X{}&$\ket{\psi_+}$&\includegraphics[width=1cm]{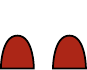}&\includegraphics[width=1cm]{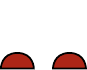}\\[\smallskipamount]
\end{tabular}
\caption{\label{fig.states}Encoding of the states sent by Alice.}
\end{figure}

The protocol we present is based on time-bin encoding with decoy state method.
The general form for a qubit is: 
\begin{equation}
\ket{\psi} = c_0 \ket{\psi_0} + c_1 e^{i \phi}\ket{\psi_1},
\end{equation}
with $c_0^2 + c_1^2 = 1$.
Time-bin qubits are composed of two temporal modes denoted as early (E) and late (L). 
We approximate the time-bin qubits using weak coherent states $\alpha$ as follows: $\ket{\psi_0} = \ket{\alpha}_{\rm{E}}\ket{0}_{\rm{L}}$ and $\ket{\psi_1} = \ket{0}_{\rm{E}}\ket{\alpha}_{\rm{L}}$.
Many QKD implementations use four BB84 states of the form $\ket{\psi} = \frac{1}{\sqrt{2}} (\ket{\psi_0} + e^{i \phi}\ket{\psi_1})$, with $\phi \in \{0,\pi/2,\pi,3\pi/2\}$ (see e.g.~\cite{Marand1995, Yuan2008} for time-bin implementations).
Here we employ only three quantum states~\cite{Fung2006, Tamaki2014, Mizutani2015} with the encoding illustrated in figure~\ref{fig.states}.
In the \Z{} basis, which is used to generate the raw key, Alice encodes the bits 0 or 1 by preparing the states $\ket{\psi_0}$ or $\ket{\psi_1}$, respectively.
The \X{} basis is used to estimate the eavesdropper's information.
In this basis, Alice sends $\ket{\psi_+} = \frac{1}{\sqrt{2}}(\ket{\psi_0} + \ket{\psi_1})$.
The probabilities to choose the bases \Z{} and \X{} are $\paz$ and $\pax$, respectively.

As we use weak coherent pulses, we prevent photon number splitting attacks~\cite{Huttner1995, Brassard2000} by using the decoy state method~\cite{Lo2005}.
We implement it using only two different mean photon numbers  $\mu_1$ and $\mu_2$, also named signal and decoy, respectively.
This one-decoy state method has recently been shown to be optimal for almost all experimental settings~\cite{Rusca2018}. 
The mean photon number $\mu = |\alpha|^2$ of each qubit is chosen at random among $\mu_1$ and $\mu_2$ with corresponding probabilities $p_1$ and $p_2$.

Bob measures the qubits either in \Z{} or \X{} basis with probabilities $\pbz$ and $\pbx$, respectively.
The \Z{} basis measurement is a direct measurement of the arrival time of the photon that allows Bob to recover the bit value.
In the \X{} basis, the coherence between two consecutive pulses is measured via an unbalanced interferometer.

\begin{figure*}
\includegraphics[width = 1.9\columnwidth]{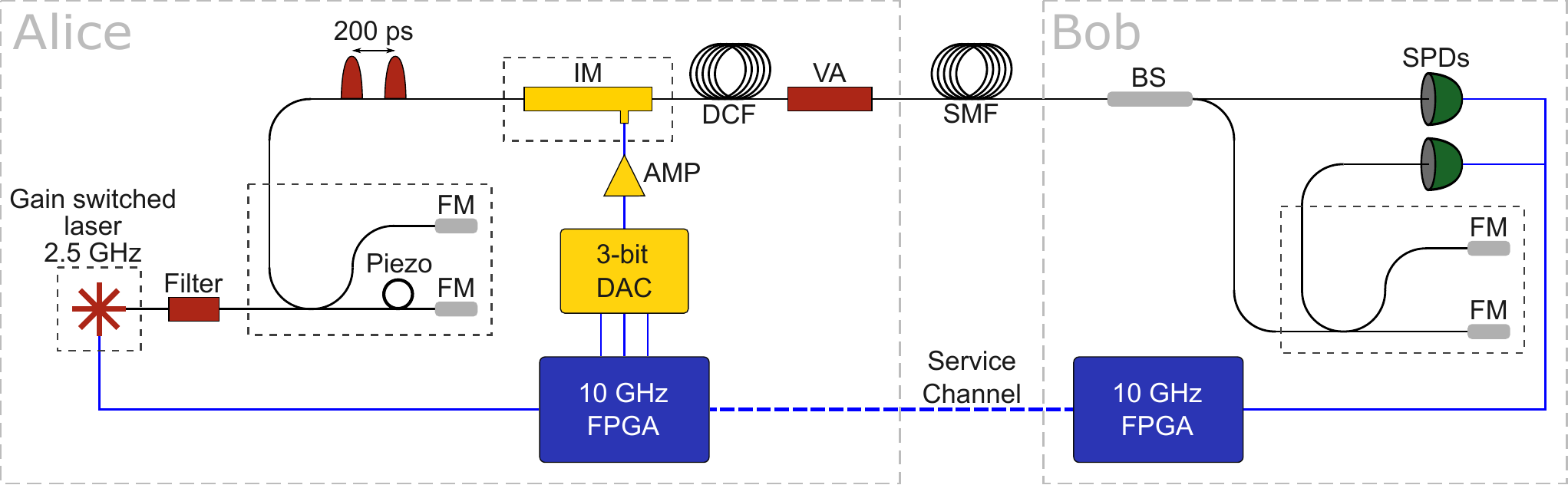}
\caption{\label{fig.implementation}Schematics of the experimental setup. Filter: 270\,pm bandpass filter; Piezo: piezoelectric fiber stretcher; FM: Faraday mirror; IM: intensity modulator; AMP: RF amplifier; DCF: dispersion compensating fiber; VA: variable attenuator; SMF: single-mode fiber; BS: beamsplitter; SPD: single-photon detector. Dashed lines represent temperature stabilized boxes.}
\end{figure*}

The experimental setup is depicted in figure~\ref{fig.implementation}.
Alice's and Bob's devices are each controlled by a field programmable gate array (FPGA) [Xilinx Kintex-7 development board] with 10\,GHz sampling rate. They are synchronized via an optical service channel based on 10\,GHz small form-factor pluggable (SFP) transceivers.
The quantum channel is composed of spools of single-mode fiber (SMF) with an attenuation of about 0.2\,dB/km.

On Alice's side, we use a fast gain-switched distributed feedback laser at 1550\,nm (Gooch \& Housego AA0701, ITU channel 33) and verify that the pulses generated at a rate of 2.5\,GHz have a random phase~\cite{Kobayashi2014}.
A narrowband filter (270\,pm) limits the spectrum of the optical pulses and thereby the effect of chromatic dispersion of the fiber link.
Additionally, since these pulses are chirped, the filtering reduces their duration to about 30\,ps~\cite{Nakazawa1990}. 
The pulses then pass through an unbalanced Michelson interferometer with an arm length difference of 200\,ps.
Its effect is to split each pulse into two pulses that are coherent with each other.
In one arm, the fiber is wrapped around a piezoelectric cylinder that is used to adjust the interferometer phase.

Our way of encoding three states has the practical advantage that it can be implemented with only one intensity modulator (IM), whereas other implementations of phase-encoding with decoy states require a phase modulator and an IM.
Moreover, the phase $\phi$ does not have to be modulated at high speed, which reduces the preparation error of the $\ket{\psi_+}$ state.

The qubit state and the pulse amplitude to encode are chosen at random by the FPGA (using a pseudo-random number generator).
Three high-speed outputs from the FPGA are linked to a three-bit programmable digital-to-analog converter (DAC) that generates radio-frequency (RF) pulses with the appropriate amplitudes (that are further amplified by a RF amplifier).
These pulses drive a lithium niobate (LiNbO$_3$) IM [IXblue] which modulates the intensity of the pulses exiting the interferometer.
Our DAC allows us to generate only four independently adjustable levels.
They correspond to the mean photon numbers $\mu_1$, $\mu_2 = \mu_1/2$, $\mu_2/2$ and $0$, after the final attenuation.
Thus in the choice of $\mu_1$ and $\mu_2$ we have the constraint $\mu_2 = \mu_1/2$.
This is not a problem since this ratio is close to the optimum value for almost all distances (see Ref.~\cite{Rusca2018}).

We pre-compensate the chromatic dispersion of the quantum channel with dispersion compensating fiber (DCF) fabricated by Corning. Its dispersion is $-130$\,ps/km/nm (@\,1550\,nm), i.e. that 1\,km of DCF compensates the dispersion of about 7.5\,km of standard SMF.
Without this DCF, already after 50\,km the overlap between two consecutive pulses at Bob's side would dramatically increase the quantum bit error rate (QBER).
Note that the the DCF is part of Alice's setup.
Thus, it does not add to the attenuation of the quantum channel.

Finally, a variable attenuator placed at the output of Alice attenuates the signal in order to set the desired mean photon number of the outgoing pulses.

\begin{figure*}
\includegraphics[width = 1.95\columnwidth]{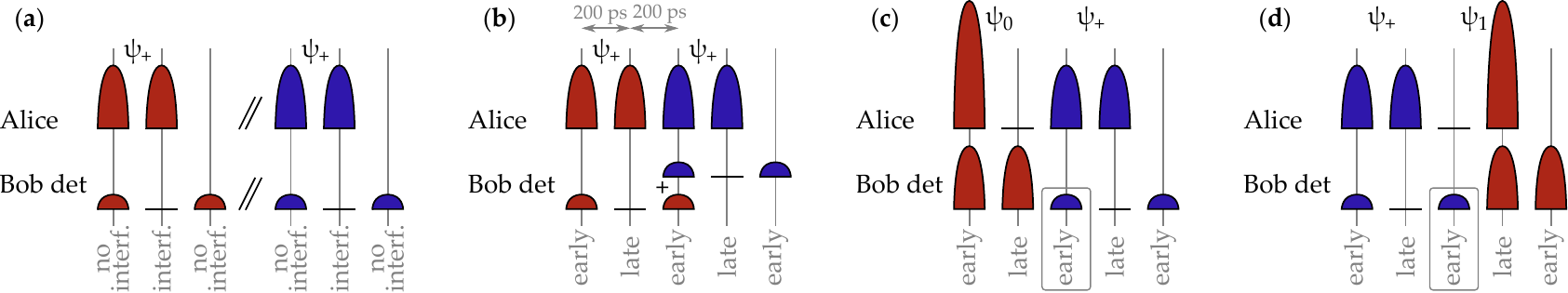}
\caption{\label{fig.Xdet} Example of the detection of two consecutive qubits in the \X{} basis. (\textbf{a}) 
In a general scheme: Alice sends the $\ket{\psi_+}$ state.
Bob can obtain detections at three different times. Interference between the two pulses is obtained in the central peak.
In the  port where the detector (Bob det) is placed, the phase is set to have destructive interference.
A second detector monitoring the light exiting the input port of the interferometer (by means of a circulator) would be necessary to detect the constructive interference peak.
(\textbf{b}) In our implementation the side peaks of two consecutive qubits overlap.
We use the time-bin \textit{early} to estimate the total number of detections in the \X{} basis when it is not affected by contributions of other qubits. This condition is fulfilled for the qubits where Alice has sent the state $\ket{\psi_+}$ preceded by $\ket{\psi_0}$ (\textbf{c}), or when she has sent $\ket{\psi_1}$ preceded by $\ket{\psi_+}$ (\textbf{d}).}
\end{figure*}

On Bob's side, the measurement basis choice is made passively via a beamsplitter with a splitting ratio of 90:10 between the \Z{} and \X{} bases. For simplicity, this splitting ratio is fixed for all measurements and we didn't optimize it for each transmission distance. In the \Z{} basis, the states are sent directly to a single-photon detector that measures the arrival time of the photon.
The measurement outcomes are either $\ket{\psi_0}$ or $\ket{\psi_1}$.
In the \X{} basis, there is an unbalanced Michelson interferometer with the same delay as the one on Alice's side.
After this second interferometer we find three pulses: a central interfering and two side peaks (see figure~\ref{fig.Xdet}a).
Note that since the interferometer delay is exactly half of the clock period, the side peaks of neighboring qubits overlap at the output port (see figure~\ref{fig.Xdet}b). 
We fix the phase difference between Alice's and Bob's interferometers such that a detection in the central interfering time-bin corresponds to the state $\ket{\psi_-} =  \frac{1}{\sqrt{2}}(\ket{\psi_0} - \ket{\psi_1})$. The photons projected onto the state $\ket{\psi_+} =  \frac{1}{\sqrt{2}}(\ket{\psi_0} + \ket{\psi_1})$ exit the interferometer via the second port, that is to say the input port. We do not detect these events.
The phase difference is kept constant with a feedback loop that takes the QBER in the \X{} basis as an error signal.
Moreover, to compensate the length fluctuations of the quantum channel, an automatic feedback loop constantly adjusts electrical delays placed between the detectors and the FPGA.
Since the sampling by the FPGA is performed at 10\,GHz, the detection bins corresponding to $\ket{\psi_0}$ and $\ket{\psi_1}$ are separated by a an "empty" time-bin whose detections can be used for this temporal tracking.

For the experimental results presented here, we use two in-house-made single-photon detectors based on InGaAs/InP negative feedback avalanche photodiodes cooled by a free-piston Stirling cooler.
They feature free-running operation, dark count rates (DCR) as low as 1\,cps~\cite{Korzh2014}, jitter as low as 50\,ps~\cite{Amri2016} and efficiencies up to 30\,\%, depending on the bias voltage and operation temperature. 

After sifting, we perform error correction in real time with the Cascade algorithm \cite{Martinez2015} on blocks of 8192\,bits.
The QBER in the \Z{} basis $\QZ$ is directly provided by the algorithm.
Privacy amplification is performed every 1000 blocks of error correction, meaning that the privacy amplification block size is $8.192 \cdot 10^6$ bits.
We determine the secret key fraction following the calculation described in~\cite{Rusca2018} using the one-decoy protocol with finite-key analysis.
The length of the extractable key is given by:
\begin{align}\label{eq:skl}
l \leq &  \sz{0} + \sz{1}(1-h(\phiz)) - \lambda_\text{EC} \nonumber\\
& - 6\log_2(19/\epsilon_\text{sec}) - \log_2(2/\epsilon_\text{cor}),
\end{align}
where $\sz{0}$ and $\sz{1}$ are the lower bound on the number of vacuum and single-photon detections in the \Z{} basis, $\phiz$ is the upper bound on the phase error rate, $\lambda_\text{EC}$ is the total number of bits revealed during the error correction, and $\epsilon_\text{sec} = 10^{-9}$ and $\epsilon_\text{cor} = 10^{-9}$ are the secrecy and correctness parameters, respectively.

If one uses two detectors in the \X{} basis monitoring both outputs of the interferometer, the estimation of $\phiz$ is straightforward, knowing the detections corresponding both to $\ket{\psi_-}$ and $\ket{\psi_+}$.
Here, we do not have direct access to $\ket{\psi_+}$.
Therefore, we use the detections in the non-interfering peaks of the \X{} basis to estimate  the total number of detections in the \X{} basis $n_{\mathsf{X}}$.
Furthermore, we have to restrict ourselves to the conclusive events $n_{\mathsf{X}\mathrm{ side}}$, which correspond to the detections in the time-bin \textit{early} when Alice has sent $\ket{\psi_+}$ preceded by $\ket{\psi_0}$, or $\ket{\psi_1}$ preceded by $\ket{\psi_+}$ (see figures~\ref{fig.Xdet}c and~\ref{fig.Xdet}d).
Using the fact that the probability to send $\ket{\psi_0}$ and the probability to send $\ket{\psi_1}$ are both equal to $\frac{1}{2} \paz$, we have
\begin{equation}
n_{\mathsf{X}} = \frac{n_{\mathsf{X}\mathrm{side}}}{\frac{1}{4}\paz}.
\end{equation} 
The extra $1/2$ term comes from the fact that half of the events exit through the input port of the interferometer.


\begin{table*}[t]
\begin{tabular}{C{1.2cm}C{1.2cm}|C{1.2cm}C{1.2cm}|C{1.2cm}C{1.2cm}C{1.6cm}|C{1.2cm}C{1.2cm}C{1.6cm}C{1.6cm}}
Length&Attn&$T$&$\tau_{\mathsf{Z}}$&$\mu_1$&$\mu_2$&block time&$\QZ$&$\phiz$&RKR&SKR\\
(km)&(dB)&(K)&($\mu$s)&&&(s)&(\%)&(\%)&(bps)&(bps)\\
\hline
101.6&20.2&203&15.7&0.06 & 0.03 & 232 & 3.67 & 2.03 & $3.53 \cdot 10^{4}$ & $1.42 \cdot 10^{4}$\\
151.6&30.2&183&19.3&0.25 & 0.11 & 360 & 3.21 & 2.12 & $2.28 \cdot 10^{4}$ & $7.21 \cdot 10^{3}$\\
202.1&40.2&183&27.0&0.39 & 0.18 & 1008 & 3.08 & 3.59 & $8.13 \cdot 10^{3}$ & $1.59 \cdot 10^{3}$\\
\end{tabular}
\caption{\label{tab.Results} Overview of experimental parameters and performance for different fiber lengths.}
\end{table*}

We tested our system over different lengths of standard SMF: 100, 150 and 200\,km.
We can adjust the temperature $T$ and the dead time $\tau$ of the detectors in order to maximize the SKR.
Lowering the temperature reduces the DCR, but also increases the after-pulse probability.
(In our setup, both detectors are placed in the same Stirling cooler meaning that they are necessarily both at the same temperature.)
Reducing the dead time increases the maximum count rates, but also increases the after-pulse probability, hence the QBER. 
At short distances, it is favourable to use a short dead time and higher temperatures above 200\,K. If we still saturate the detector it's advantageous to reduce $\mu_1,\mu_2$.
In the \X{} basis, the detector is never saturated, hence its dead time was kept constant at 75.8\,$\mu$s for all measurements.
At longer distances we lower the temperature in order to reduce the DCR and can afford longer dead times in order to limit the contribution of the after-pulses without affecting the detection rate.
For all distances the parameters for the state preparation were $\paz=0.9$ ($\pax=0.1$) and $p_1 = 0.6$ ($p_2 = 0.4$).
The experimental parameters, the detector settings and the performances achieved at each distance are summarized in table~\ref{tab.Results}.

\begin{figure}[b]
\includegraphics[trim={0.2cm 0 1.3cm 0},clip,width = \columnwidth]{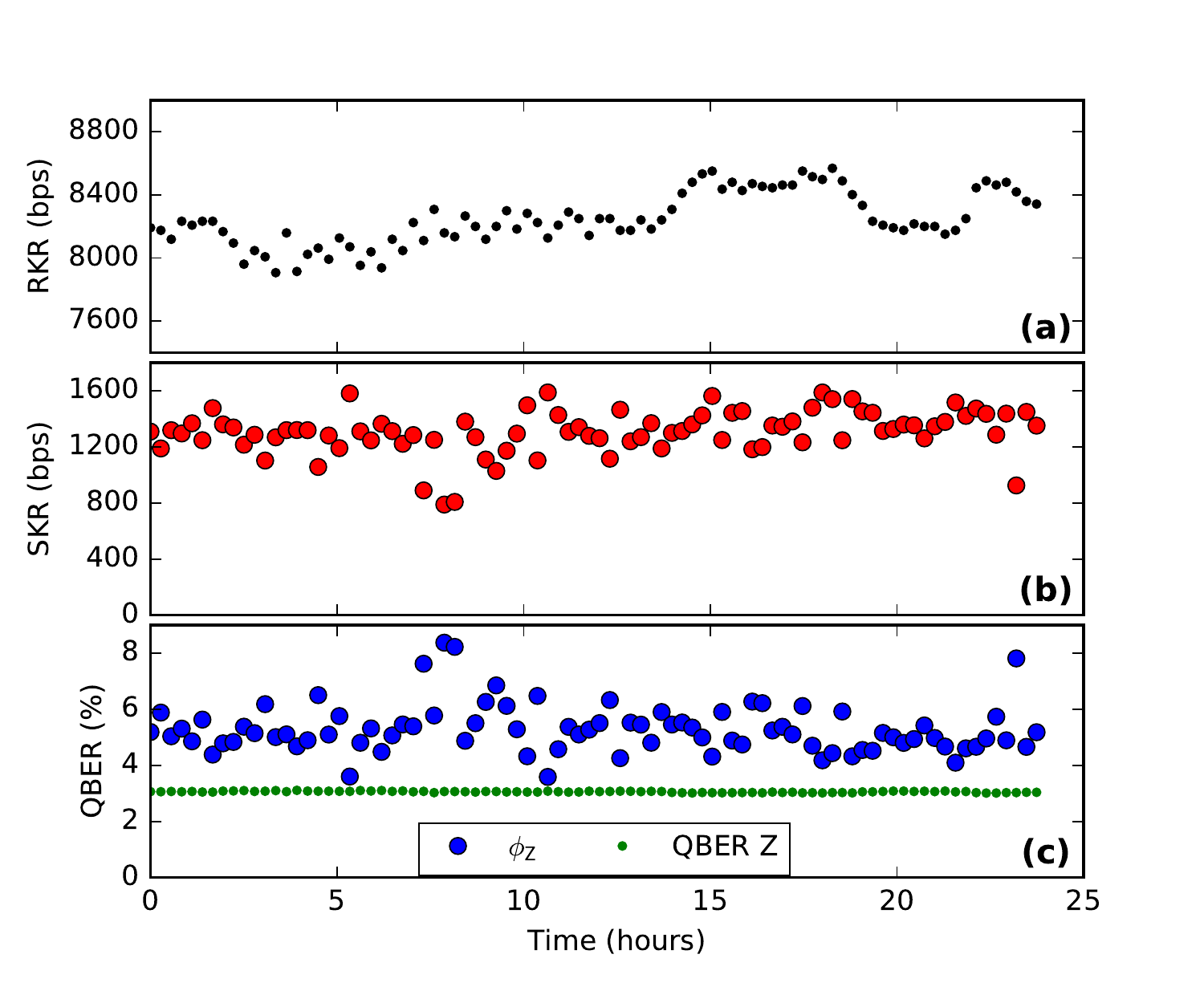}
\caption{\label{fig.stab} System stability over 24\,hours with 200\,km of optical fiber. (\textbf{a}) RKR, (\textbf{b}) SKR, and (\textbf{c}) corresponding $\QZ$ and $\phiz$ as a function of the time.}
\end{figure}

We have also tested the stability of the system over a continuous period of 24\,hours at a distance of 200\,km.
The phase stabilization of the two interferometers and the temporal alignment was automatic with no interruption.
The raw key rate (RKR), the SKR, $\QZ$ and $\phiz$  are shown in figure~\ref{fig.stab} as a function of time.
In total, 87\,blocks of privacy amplification were acquired, corresponding to 712\,Mb of raw key, from which 114\,Mb of secret key were extracted.


The contributions to the QBER are the detection time uncertainty, the limited extinction ratio in the state preparation, and the DCR and after-pulses of the detectors.
In the \X{} basis, the phase fluctuations of the interferometers are an additional source of QBER.

We estimate that the 3\,\% of QBER (@\,200\,km) in the \Z{} basis can be roughly accounted for as follows: 1.5\,\% is due to the timing jitter (obtained by a comparison with a measurement using a low jitter superconducting single-photon detector). 1\,\% is due to imperfect intensity modulation.
Statically the extinction ratio of the IM is $-30$\,dB, but is reduced at 5\,GHz modulation frequency.
Moreover, our RF amplifier is AC-coupled, which implies that the amplitude of an electrical pulse depends to some extent on the sequence preceding it.
In fact, this is an additional advantage of using the one-decoy instead of the two-decoy method, since adding a third $\mu$ would increase the uncertainty in the pulse amplitudes due to the AC coupling.
Finally, 0.5\,\% is due to detector noise and afterpulsing.
In the \X{} basis, the QBER is less affected by pulse amplitude fluctuations.

At short distance, $\QZ$ is higher than $\phiz$ because we use a shorter dead time in the \Z{} basis in order to obtain higher detection rates, and hence we increase the contribution of the after-pulses.
Moreover, the detection timing jitter is increased if the detector saturates.
Variations of the quantum channel length which are not perfectly compensated by the automatic stabilization, lead to reduced RKR and consequently to higher QBER.
The contribution of the DCR to the QBER is significant only in the \X{} basis, due to the smaller count rates.
It accounts for about 40\,\% of the errors at 200\,km.
Finally, Alice's and Bob's interferometers feature intrinsic visibilities exceeding 99.5\,\%. However, phase fluctuations reduce the total visibility and and affect $\phiz$.


In conclusion, we implemented a state-of-the-art QKD system featuring a 2.5\,GHz clock rate. Our system is based on a protocol using three quantum states with time-bin encoding and the one-decoy state method.
We performed a secret key exchange over 200\,km of standard SMF with a rate of about 1.5\,kbps, which is a one order of magnitude improvement with respect to previous results~\cite{Korzh2015, Frohlich2017}.


\section*{Acknowledgements}
We would like to acknowledge Jes\'us Mart\'inez-Mateo for providing the error correction code and Charles Ci Wen Lim for useful discussions.
We thank the Swiss NCCR QSIT and Davide Rusca thanks the EUs H2020 programme under the Marie Sk\l{}odowska-Curie project QCALL (GA 675662)  for financial support.

\bibliography{bib_StallionInGaAs}
\end{document}